`\input amstex
\documentstyle{amsppt}
%
\catcode`@=11
\redefine\output@{%
  \def\break{\penalty-\@M}\let\par\endgraf
  \ifodd\pageno\global\hoffset=105pt\else\global\hoffset=8pt\fi  
  \shipout\vbox{%
    \ifplain@
      \let\makeheadline\relax \let\makefootline\relax
    \else
      \iffirstpage@ \global\firstpage@false
        \let\rightheadline\frheadline
        \let\leftheadline\flheadline
      \else
        \ifrunheads@ 
        \else \let\makeheadline\relax
        \fi
      \fi
    \fi
    \makeheadline \pagebody \makefootline}%
  \advancepageno \ifnum\outputpenalty>-\@MM\else\dosupereject\fi
}
\catcode`\@=\active
\nopagenumbers
\def\negskp{\hskip -2pt}
\def\rot{\operatorname{rot}}
\def\const{\operatorname{const}}
\def\grad{\operatorname{grad}}
\def\divr{\operatorname{div}}
\def\blue#1{#1}
\catcode`#=11\def\diez{#}\catcode`#=6
\catcode`_=11\def\podcherkivanie{_}\catcode`_=8
\def\mycite#1{\cite{\blue{#1}}}
\def\mytag#1{%
    \tag#1}
\def\mythetag#1{\thetag{\blue{#1}}}
\def\myrefno#1{\no#1}
\def\myhref#1#2{\blue{#2}}
\def\myEarXivlink{\myhref{http://arXiv.org}{http:/\negskp/arXiv.org}}
\pagewidth{360pt}
\pageheight{606pt}
\topmatter
\title
Burgers space versus real space\\
\lowercase{in the nonlinear theory of dislocations.}
\endtitle
\rightheadtext{Burgers space versus real space \dots}
\author
Ruslan Sharipov
\endauthor
\address Rabochaya street 5, 450003 Ufa, Russia
\endaddress
\email \vtop to 30pt{\hsize=280pt\noindent
\myhref{mailto:R\podcherkivanie Sharipov\@ic.bashedu.ru}
{R\_\hskip 1pt Sharipov\@ic.bashedu.ru}\newline
\myhref{mailto:r-sharipov\@mail.ru}
{r-sharipov\@mail.ru}\newline
\myhref{mailto:ra\podcherkivanie sharipov\@lycos.com}{ra\_\hskip 1pt
sharipov\@lycos.com}\vss}
\endemail
\urladdr
\myhref{http://www.geocities.com/r-sharipov}
{http:/\negskp/www.geocities.com/r-sharipov}
\endurladdr
\abstract
    Some double space tensorial quantities in the nonlinear
theory of dislocations are considered. Their real space counterparts
are introduced.
\endabstract
\endtopmatter
\loadbold
\TagsOnRight
\document

\head
1. Introduction.
\endhead
    The Burgers vector is a quantitative characteristic of each
dislocation line in a crystal. It is similar to the charge or the
mass of elementary particles, but it is a vectorial quantity. The
Burgers space $\Bbb B$ is introduced as a container for all of the 
Burgers vectors (see \mycite{1}). The Burgers space is an imaginary 
space, it is different from the real space $\Bbb E$ where the actual
evolution of a crystal occurs. The Burgers space can be understood 
as a copy of the real space filled with the infinite ideal (non-distorted)
crystalline grid that does not move and does not evolve in any other
way. Like the real space, it is equipped with the metric given by
the metric tensor and the dual metric tensor. Their components
$\overset\sssize\star\to g\vphantom{g}_{ij}$ and $\overset\sssize\star
\to g\vphantom{g}^{ij}$ are related to some Cartesian coordinates $x^1,\,x^2,\,x^3$ in $\Bbb B$. The components of the metric tensor and
the dual metric tensor in the real space $\Bbb E$ are denoted by
$g_{pq}$ and $g^{pq}$, i\.\,e\. without the star. For the sake of
simplicity, below we assume $g_{pq}$ and $g^{pq}$ to be related to some
Cartesian coordinates $y^1,\,y^2,\,y^3$ in $\Bbb E$ (though we could
choose curvilinear coordinates either). Due to the choice of Cartesian coordinates the components of metric tensors are constants (see 
\mycite{2}).\par
    The basic differential equations describing the kinematics of
a dislocated crystal are written in terms of the tensor fields
$\hat\bold T$, $\bold T$, $\boldsymbol\rho$, $\bold j$\,, and 
$\bold w$ (see \mycite{1}):
$$
\gather
\hskip -2em
\frac{\partial\hat T^i_k}{\partial t}+j^{\,i}_{\,k}
=-\nabla_{\!k}w^i,
\mytag{1.1}\\
\hskip -2em
\frac{\partial T^i_k}{\partial t}
=-\sum^3_{p=1}\nabla_{\!k}(v^p\ T^i_p),
\mytag{1.2}\\
\hskip -2em
w^i=\sum^3_{p=1}v^p\ \hat T^i_p,
\mytag{1.3}\\
\hskip -2em
\sum^3_{q=1}\sum^3_{p=1}
\sum^3_{r=1}\sum^3_{m=1}\omega_{kqp}\ g^{qr}\,g^{pm}\,\nabla_{\!r}
\hat T^{\,i}_m=\rho^{\,i}_k.
\mytag{1.4}
\endgather
$$
The third equation \mythetag{1.3} is based on the conjecture~4.1 suggested
in \mycite{3}. If this conjecture appears to be not valid, then
\mythetag{1.3} will be replaced by some other equation. However, this
will not affect the methods we use in the present paper. With this 
explicit reservation, we continue assuming below that the conjecture
from \mycite{3} is valid and that \mythetag{1.3} is a true expression 
for $w^i$.\par
    The tensor fields $\hat\bold T$, $\bold T$, $\boldsymbol\rho$, 
$\bold j$\,, and $\bold w$ are double space tensor fields. Their upper 
index $i$ in \mythetag{1.1}, \mythetag{1.2}, \mythetag{1.3}, and \mythetag{1.4} is associated with the Burgers space $\Bbb B$, while 
their components are functions of the coordinates $y^1,\,y^2,\,y^3$ in
the real space $\Bbb E$:
$$
\align
&\hat T^i_k=\hat T^i_k(t,y^1,y^2,y^3),\\
&T^i_k=T^i_k(t,y^1,y^2,y^3),\\
&j^{\,i}_{\,k}=j^{\,i}_{\,k}(t,y^1,y^2,y^3),
\mytag{1.5}\\
&\rho^{\,i}_k=\rho^{\,i}_k(t,y^1,y^2,y^3),\\
&w^i=w^i(t,y^1,y^2,y^3).
\endalign
$$
Note that $v^1,\,v^2,\,v^3$ are the components of the velocity vector
$\bold v$ describing the motion of a medium. Unlike \mythetag{1.5},
they are components of a real space tensor:
$$
\hskip -2em
v^p=v^p(t,y^1,y^2,y^3).
\mytag{1.6}
$$
The index $p$ in \mythetag{1.2}, \mythetag{1.3}, and in \mythetag{1.6} 
is associated with the real space. Apart from \mythetag{1.5} and \mythetag{1.6}, in \mythetag{1.4} we see the components of the so-called
{\it volume tensor}, they are denoted $\omega_{kqp}$. Due to the choice 
of the Cartesian coordinates $y^1,\,y^2,\,y^3$ in $\Bbb E$ they are
constants: $\omega_{kqp}=\const$.\par
     The concept of the Burgers space and the double space tensor fields
associated with it are convenient tools in understanding the microscopic
structure of dislocations (in terms of a crystalline grid and interatomic
bonds). They are also convenient in deriving the basic differential
equations \mythetag{1.1}, \mythetag{1.2}, and \mythetag{1.4}. However,
macroscopically, e\.\,g\. in explaining real stress-strain curves or in
computer simulation of the behavior of real crystals, it would be better
to deal with purely real space tensor fields. The main goal of the present
paper is to find some purely real space substitutes for the tensor fields
\mythetag{1.5} and derive the differential equations for them in place of
\mythetag{1.1}, \mythetag{1.2}, and \mythetag{1.4}. In part, this work
is already done in \mycite{1}. Below we complete this work and thus lay
a foundation for the further development of our approach to the theory
of dislocations.\par
\head
2. The elastic and plastic deformation tensors.
\endhead
     In \mycite{1} the {\it elastic\/} and the {\it plastic\/} deformation
tensors $\hat\bold G$ and $\check\bold G$ for a dislocated crystalline
matter were suggested. They are given by the formulas
$$
\gather
\hskip -2em
\hat G_{pq}=\sum^3_{i=1}\sum^3_{j=1}\overset\sssize\star\to
g\vphantom{g}_{ij}\ \hat T^i_p\ \hat T^j_q,
\mytag{2.1}\\
\hskip -2em
\check G^p_q=\sum^3_{i=1}\hat S^p_i\ T^i_q,
\mytag{2.2}
\endgather
$$
where $\hat S^p_i$ are the components of the inverse matrix $\hat\bold S=
\hat\bold T^{-1}$. The {\it total\/} deformation tensor $\bold G$ is 
defined by the similar formula
$$
\hskip -2em
G_{pq}=\sum^3_{i=1}\sum^3_{j=1}\overset\sssize\star\to
g\vphantom{g}_{ij}\ T^i_p\ T^j_q.
\mytag{2.3}
$$
From \mythetag{2.1}, \mythetag{2.2}, and \mythetag{2.3} one easily
derives the equality
$$
\hskip -2em
G_{pq}=\sum^3_{r=1}\sum^3_{s=1}\check G^{\,r}_p\ \hat G_{rs}\ 
\check G^{\,s}_q.
\mytag{2.4}
$$
The equality \mythetag{2.4} is known as the {\it multiplicative decomposition}
of the total deformation tensor into the elastic and plastic parts. This
decomposition was first suggested in \mycite{4} for the theory of plastic
glassy media.\par
     Both deformation tensors $\hat\bold G$ and $\check\bold G$ are purely
real space tensors, and so is the total deformation tensor $\bold G$.
The tensor $\hat\bold G$ is considered as a real space substitute for
$\hat\bold T$, and $\bold G$ is such a substitute for $\bold T$. Passing
from $\hat\bold T$ to $\hat\bold G$, we loose a part of information
contained in $\hat\bold T$. However, this is that very part of information which is inessential. Indeed, if $\bold O$ is a rotation matrix in the Burgers space, i\.\,e\. if
$$
\sum^3_{i=1}\sum^3_{j=1}\overset\sssize\star\to
g\vphantom{g}_{ij}\ O^{\,i}_p\ O^{\,j}_q=\overset\sssize\star\to
g\vphantom{g}_{pq},
$$
then the distorsion tensor $\hat\bold T'=\bold O\cdot\hat\bold T$ represent
the same extent of deformation in interatomic bonds as the the tensor 
$\hat\bold T$. Similarly, a crystalline body with the distorsion field
$\bold T'=\bold O\cdot\bold T$ has the same shape as if it had the 
distorsion field $\bold T$ (the matrix $\bold O$ is assumed to be a constant
rotation matrix).\par
     The deformation tensors $\bold G$ and $\hat\bold G$ satisfy the
following partial differential equations describing their time evolution:
$$
\gather
\hskip -2em
\frac{\partial G_{pq}}{\partial t}+\sum^3_{r=1}v^r\,\nabla_{\!r}G_{pq}=
-\sum^3_{r=1}G_{rq}\,\nabla_{\!p}v^r-\sum^3_{r=1}G_{pr}\,\nabla_{\!q}v^r,
\mytag{2.5}\\
\vspace{2ex}
\hskip -2em
\gathered
\frac{\partial\hat G_{pq}}{\partial t}+\sum^3_{r=1}v^r\,
\nabla_{\!r}\hat G_{pq}=-\sum^3_{r=1}\nabla_{\!p}v^r\,
\hat G_{rq}-\sum^3_{r=1}\hat G_{pr}\,\nabla_{\!q}v^r+\\
+\sum^3_{r=1}\theta^{\,r}_p\,\hat G_{rq}+\sum^3_{r=1}
\hat G_{pr}\,\theta^{\,r}_q.
\endgathered
\mytag{2.6}
\endgather
$$
Here $\theta^{\,r}_p$ and $\theta^{\,r}_q$ are the components of another
one purely real space tensor field $\boldsymbol\theta$. They are given
by the following formula:
$$
\hskip -2em
\theta^{\,r}_q=\nabla_{\!q}v^r-\sum^3_{i=1}\hat S^r_i\,j^{\,i}_q
-\sum^3_{i=1}\hat S^r_i\,\nabla_{\!q}w^i
+\sum^3_{i=1}\sum^3_{p=1}v^p\,\hat S^r_i\,\nabla_{\!p}\hat T^i_q.
\mytag{2.7}
$$
The differential equation \mythetag{2.5} is derived from \mythetag{1.2},
while \mythetag{2.6} is derived from \mythetag{1.1} (see \mycite{1} for
details). The components of the plastic deformation tensor \mythetag{2.2}
satisfy the following differential equation similar to \mythetag{2.6}:
$$
\hskip -2em
\frac{\partial\check G^{\,p}_q}{\partial t}+
\sum^3_{r=1}v^r\,\nabla_{\!r}\check G^{\,p}_q=
\sum^3_{r=1}\left(\check G^{\,r}_q\,\nabla_{\!r}v^p
-\nabla_{\!q}v^r\,\check G^{\,p}_r\right)-\sum^3_{r=1}
\theta^{\,p}_r\,\check G^{\,r}_q.
\mytag{2.8}
$$
In deriving \mythetag{2.8} both equations \mythetag{1.1} and 
\mythetag{1.2} are used (see \mycite{1}).\par
     Note that the equations \mythetag{2.6} and \mythetag{2.8}
do coincide with the corresponding equations in the theory of
amorphous materials (see \mycite{4} where these equations were 
written for the first time). The tensorial parameter $\boldsymbol
\theta$ is proposed as a new parameter of media like the density,
the specific heat capacity, the dielectric permittivity, the elastic
moduli, etc (see \mycite{5} where some experimental schemes for 
measuring this parameter $\boldsymbol\theta$ are discussed in brief).
According to \mycite{5}, the parameter $\boldsymbol\theta$ is
interpreted as a tensor that determines the {\it stress relaxation 
rate}.\par
     Looking at \mythetag{2.2} and \mythetag{2.7}, we see that 
the components of the inverse matrix $\hat\bold S=\hat\bold T^{-1}$
are used for to transform the index $i$, which is associated 
with the Burgers space, into the real space indices $p$ and $r$ 
respectively. By analogy, we can use $\hat S^p_i$ in order to define
the pair of purely real space tensor fields $\bold R$ and $\bold J$:
$$
\xalignat 2
&\hskip -2em
R^p_q=\sum^3_{i=1}\hat S^p_i\ \rho^{\,i}_q,
&&J^p_q=\sum^3_{i=1}\hat S^p_i\ j^{\,i}_{\,q}.
\quad
\mytag{2.9}
\endxalignat
$$
Like $\boldsymbol\rho$ and $\bold j$\,, the tensor fields \mythetag{2.9}
characterize the density and the flow of dislocations respectively.
Now, under the assumption that the conjecture~4.1 from \mycite{3} is valid,
if we substitute \mythetag{1.3} into \mythetag{2.7}, we derive
$$
\hskip -2em
\theta^{\,r}_q=-J^{\,r}_q+\sum^3_{i=1}\sum^3_{p=1}v^p\,
\hat S^r_i\,(\nabla_{\!p}\hat T^i_q-\nabla_{\!q}\hat T^i_p).
\mytag{2.10}
$$
In order to exclude the double space tensors $\hat\bold S$ and
$\hat\bold T$ from \mythetag{2.10} we introduce the tensor field 
$\hat\bold R$ with the following components:
$$
\hskip -2em
\hat R^{\,r}_{pq}=
\sum^3_{i=1}\hat S^r_i\,(\nabla_{\!p}
\hat T^i_q-\nabla_{\!q}\hat T^i_p).
\mytag{2.11}
$$
The tensor field $\hat\bold R$ is closely related to the tensor field
$\bold R$ with the components \mythetag{2.9}. Indeed, taking into account
the well-known identity
$$
\sum^3_{s=1}\omega_{slc}\ \omega^{spq}=
\delta^p_l\,\delta^q_c-\delta^p_c\,\delta^q_l,
$$
from the equalities \mythetag{2.11} and \mythetag{1.4} we derive
$$
\allowdisplaybreaks
\gather
\hat R^{\,r}_{pq}=\sum^3_{i=1}\hat S^r_i\,(\nabla_{\!p}
\hat T^i_q-\nabla_{\!q}\hat T^i_p)=\sum^3_{i=1}\sum^3_{m=1}\sum^3_{n=1}
\hat S^r_i\,(\delta^m_p\,\delta^n_q-\delta^m_q\,\delta^n_p)\,
\nabla_{\!m}T^i_n=\\
=\sum^3_{i=1}\sum^3_{s=1}\sum^3_{m=1}\sum^3_{n=1}\hat S^r_i\,
\omega^{smn}\,\omega_{spq}\,\nabla_{\!m}T^i_n=
\sum^3_{i=1}\sum^3_{s=1}\sum^3_{m=1}\sum^3_{n=1}\sum^3_{k=1}
\sum^3_{l=1}\sum^3_{c=1}\hat S^r_i\,\omega_{klc}\,g^{sk}\,\times\\
\times\,g^{lm}\,g^{cn}\,\omega_{spq}\,\nabla_{\!m}T^i_n=
\sum^3_{i=1}\sum^3_{s=1}\sum^3_{k=1}\hat S^r_i\,g^{sk}\,\omega_{spq}\,
\rho^{\,i}_k,
\endgather
$$
Now, if we remember \mythetag{2.9}, the result of the above
calculations can be written as
$$
\hskip -2em
\hat R^{\,r}_{pq}=\sum^3_{s=1}\sum^3_{k=1}\omega_{spq}\ g^{sk}\,R^r_k.
\mytag{2.12}
$$
By means of the other well-known identity
$$
\sum^3_{p=1}\sum^3_{q=1}\omega_{spq}\ \omega^{kpq}=
2\,\delta^k_s
$$
one can easily derive the converse equality to \mythetag{2.12}
that expresses $\bold R$ through $\hat\bold R$:
$$
R^{\,r}_k=\frac{1}{2}\sum^3_{s=1}\sum^3_{p=1}\sum^3_{q=1}g_{sk}
\ \omega^{spq}\ \hat R^r_{pq}.
$$
Applying \mythetag{2.11} to \mythetag{2.10} and taking into account
\mythetag{2.12}, we get
$$
\hskip -2em
\theta^{\,r}_q=-J^{\,r}_q+\sum^3_{s=1}\sum^3_{p=1}\sum^3_{k=1}
\omega_{spq}\ v^p\ g^{sk}\,R^{\,r}_k.
\mytag{2.13}
$$
Now let's remember the theorem~3.2 from \mycite{3}. In terms of the
tensor fields \mythetag{2.9} this theorem is formulated as follows.
\proclaim{Theorem 2.1}In the case of frozen dislocations the tensor
fields $\bold R$ and $\bold J$ are related to each other by the 
formula
$$
\hskip -2em
J^{\,r}_q=\sum^3_{s=1}\sum^3_{p=1}\sum^3_{k=1}
\omega_{spq}\ v^p\ g^{sk}\,R^{\,r}_k.
\mytag{2.14}
$$
\endproclaim
Comparing the equalities \mythetag{2.14} and \mythetag{2.13}, we derive
another theorem.
\proclaim{Theorem 2.2}The case of frozen dislocations is that very
case when $\boldsymbol\theta=0$, i\.\,e\. when no stress relaxation
occurs.
\endproclaim
    Note that the equality \mythetag{2.13} depend on the conjecture~4.1 from
\mycite{3}. If this conjecture is not valid, we could have some extra
terms in its right hand side. However, due to the theorem~3.4 from
\mycite{3}, the above theorem~2.2 holds irrespective to the conjecture~4.1
from \mycite{3}.
\head
3. Spatial derivatives of the deformation tensors.
\endhead
    In the previous section~2 we have rewritten the three basic equations
\mythetag{1.1}, \mythetag{1.2}, and \mythetag{1.3} in terms of the purely
real space tensor fields $\bold T$, $\hat\bold T$, $\bold R$, and $\bold J$. 
\pagebreak 
The equation \mythetag{1.1} has been transformed into \mythetag{2.5}, the equation \mythetag{1.2} --- into \mythetag{2.6}, and the equation
\mythetag{1.3} --- into \mythetag{2.13}. In deriving \mythetag{2.13} from
\mythetag{1.3} we used \mythetag{1.4}. However, the equation \mythetag{1.4}
itself is not yet transformed to the form free of double space tensors.
This problem is considered below in the present section.\par
    The equation \mythetag{1.4} contains only the spatial derivatives of
the distorsion tensor $\hat\bold T$. Therefore, it is convenient to denote
$$
\xalignat 2
&\hskip -2em
\hat Z^{\,r}_{pq}=\sum^3_{i=1}\hat S^r_i\ \nabla_{\!p}\hat T^i_q,
&&Z^{\,r}_{pq}=\sum^3_{i=1}S^r_i\ \nabla_{\!p}T^i_q.
\mytag{3.1}
\endxalignat
$$
Here $\hat S^r_i$ are the components of the inverse matrix $\hat\bold 
S=\hat\bold T^{-1}$, and $S^r_i$ are the components of the other inverse
matrix $\bold S=\bold T^{-1}$. The quantities $\hat Z^i_{pq}$ and $Z^i_{pq}$
introduced in \mythetag{3.1} are the components of two purely real space
tensor fields $\hat\bold Z$ and $\bold Z$. Despite to the similarity of
the formulas \mythetag{3.1}, the tensor fields $\hat\bold Z$ and $\bold Z$
are somewhat different. Since $\bold Z$ is produced by the compatible
distorsion tensor, it is symmetric:
$$
\hskip -2em
Z^{\,r}_{pq}=Z^{\,r}_{qp}.
\mytag{3.2}
$$
Indeed, by definition, the components of the compatible distorsion tensor  
$\bold T$ are given by partial derivatives (see \mycite{1} and \mycite{3}):
$$
\hskip -2em
T^i_q=\frac{\partial x^i}{\partial y^q}.
\mytag{3.3}
$$
From \mythetag{3.3} in Cartesian coordinates $y^1,\,y^2,\,y^3$ we
immediately get
$$
\hskip -2em
\nabla_{\!p}T^i_q=\frac{\partial^2 x^i}{\partial y^p\ \partial y^q}
=\frac{\partial^2 x^i}{\partial y^q\ \partial y^p}=\nabla_{\!q}T^i_p.
\mytag{3.4}
$$
Applying \mythetag{3.4} to the second equality \mythetag{3.1}, we find
that the above symmetry condition \mythetag{3.2} is a valid equality.
As for the tensor $\hat\bold Z$ produced from the incompatible distorsion
tensor $\hat\bold T$, its components are usually not symmetric.\par
    In the next step we apply the operator $\nabla_{\!p}=\partial/\partial
y^p$ to the components $G_{qk}$ of the deformation tensor $\bold G$. From
\mythetag{2.3}, since $\overset\sssize\star\to g\vphantom{g}_{ij}=\const$,
we derive
$$
\hskip -2em
\nabla_{\!p}G_{qk}=\sum^3_{r=1}Z^{\,r}_{pq}\,G_{rk}
+\sum^3_{r=1}Z^{\,r}_{pk}\,G_{rq}.
\mytag{3.5}
$$
By means of cyclic transposition of indices $p\to q\to k\to p$ in
\mythetag{3.5} we get
$$
\align
&\hskip -2em
\nabla_{\!q}G_{kp}=\sum^3_{r=1}Z^{\,r}_{qk}\,G_{rp}
+\sum^3_{r=1}Z^{\,r}_{qp}\,G_{rk},
\mytag{3.6}\\
&\hskip -2em
\nabla_{\!k}G_{pq}=\sum^3_{r=1}Z^{\,r}_{kp}\,G_{rq}
+\sum^3_{r=1}Z^{\,r}_{kq}\,G_{rp}.
\mytag{3.7}
\endalign
$$
By adding \mythetag{3.5} and \mythetag{3.6}, then subtracting
\mythetag{3.7} and taking into account \mythetag{3.2} we derive
the following equality for the quantities $Z^{\,r}_{pq}$:
$$
\nabla_{\!p}G_{qk}+\nabla_{\!q}G_{kp}-\nabla_{\!k}G_{pq}
=2\sum^3_{r=1}Z^{\,r}_{pq}\,G_{rk}
$$
If we denote by $[\,G^{-1}]^{kr}$ the components of the inverse
matrix $\bold G^{-1}$, we can write
$$
\hskip -2em
Z^{\,r}_{pq}=\sum^3_{k=1}\frac{
\nabla_{\!p}G_{qk}+\nabla_{\!q}G_{kp}-\nabla_{\!k}G_{pq}}{2}\
[\,G^{-1}]^{kr}.
\mytag{3.8}
$$\par
     By applying $\nabla_{\!p}=\partial/\partial y^p$ \,to the components
$\hat G_{qk}$ of the elastic deformation tensor $\hat\bold G$ one can 
derive the equality similar to \mythetag{3.5}:
$$
\hskip -2em
\nabla_{\!p}\hat G_{qk}=\sum^3_{r=1}\hat Z^{\,r}_{pq}\,\hat G_{rk}
+\sum^3_{r=1}\hat Z^{\,r}_{pk}\,\hat G_{rq}.
\mytag{3.9}
$$
However, $\hat Z^{\,r}_{pq}$ are not symmetric in $p$ and $q$. Therefore,
\mythetag{3.9} is insufficient for expressing the tensor $\hat\bold Z$
through $\hat\bold G$ and $\nabla\hat\bold G$. The number of components
$\hat Z^{\,r}_{pq}$ is equal to $27$, while \mythetag{3.9} is symmetric in $k$ and $q$. This means that in \mythetag{3.9} we have only $18$ linear algebraic equations for $\hat Z^{\,r}_{pq}$ different from each other. Nine equations are obviously lacking.\par
     Let's expand the tensor field $\hat\bold Z$ into symmetric and 
skew-symmetric parts, the skew-symmetric part being given by the tensor 
$\hat\bold R$ with the components \mythetag{2.11}:
$$
\hskip -2em
\hat\bold Z=\hat\bold H+\frac{1}{2}\,\hat\bold R.
\mytag{3.10}
$$
Due to \mythetag{3.10} we can rewrite the equation \mythetag{3.9} as follows:
$$
\hskip -2em
\gathered
\sum^3_{r=1}\hat H^{\,r}_{pq}\,\hat G_{rk}
+\sum^3_{r=1}\hat H^{\,r}_{pk}\,\hat G_{rq}=\\
=\nabla_{\!p}\hat G_{qk}-
\frac{1}{2}\sum^3_{r=1}\hat R^{\,r}_{pq}\,\hat G_{rk}
-\frac{1}{2}\sum^3_{r=1}\hat R^{\,r}_{pk}\,\hat G_{rq}.
\endgathered
\mytag{3.11}
$$
Because of the symmetry of $\hat\bold H$ we can apply to \mythetag{3.11}
the same method as used in deriving \mythetag{3.8} from \mythetag{3.5}.
As a result we obtain
$$
\hskip -2em
\gathered
\hat H^r_{pq}=\sum^3_{k=1}\frac{
\nabla_{\!p}\hat G_{qk}+\nabla_{\!q}\hat G_{kp}
-\nabla_{\!k}\hat G_{pq}}{2}\
[\,\hat G^{-1}]^{kr}-\\
-\frac{1}{2}\sum^3_{k=1}\sum^3_{s=1}\hat R^s_{pk}\,
\hat G_{sq}\,[\,\hat G^{-1}]^{kr}
-\frac{1}{2}\sum^3_{k=1}\sum^3_{s=1}\hat R^s_{qk}\,
\hat G_{sp}\,[\,\hat G^{-1}]^{kr}.
\endgathered
\mytag{3.12}
$$
The formula \mythetag{3.12} shows that we cannot express both
$\hat\bold H$ and $\hat\bold R$ through $\hat\bold G$ and 
$\nabla\hat\bold G$. However, if we take the components of
$\hat\bold R$ for independent variables, then we can express
$\bold H$ through $\hat\bold G$, $\nabla\hat\bold G$, and
$\hat\bold R$. From \mythetag{3.12} and \mythetag{3.10} we
derive
$$
\hskip -2em
\gathered
\hat Z^{\,r}_{pq}=\sum^3_{k=1}\frac{
\nabla_{\!p}\hat G_{qk}+\nabla_{\!q}\hat G_{kp}
-\nabla_{\!k}\hat G_{pq}}{2}\
[\,\hat G^{-1}]^{kr}+\frac{1}{2}\,\hat R^r_{pq}\ -\\
-\,\frac{1}{2}\sum^3_{k=1}\sum^3_{s=1}\hat R^s_{pk}\,
\hat G_{sq}\,[\,\hat G^{-1}]^{kr}
-\frac{1}{2}\sum^3_{k=1}\sum^3_{s=1}\hat R^s_{qk}\,
\hat G_{sp}\,[\,\hat G^{-1}]^{kr}.
\endgathered
\mytag{3.13}
$$
Now let's substitute \mythetag{2.12} into \mythetag{3.13}. As a result
we obtain
$$
\hskip -2em
\gathered
\hat Z^{\,r}_{pq}=\sum^3_{k=1}\frac{
\nabla_{\!p}\hat G_{qk}+\nabla_{\!q}\hat G_{kp}
-\nabla_{\!k}\hat G_{pq}}{2}\
[\,\hat G^{-1}]^{kr}\ +\\
+\ \frac{1}{2}\,\sum^3_{s=1}\sum^3_{k=1}\omega_{spq}\ 
g^{sk}\,R^r_k\ -\\
-\ \frac{1}{2}\sum^3_{k=1}\sum^3_{s=1}
\sum^3_{m=1}\sum^3_{n=1}\omega_{mpk}\ g^{mn}\,R^s_n\,
\hat G_{sq}\,[\,\hat G^{-1}]^{kr}\ -\\
-\ \frac{1}{2}\sum^3_{k=1}\sum^3_{s=1}\sum^3_{m=1}
\sum^3_{n=1}\omega_{mqk}\
g^{mn}\,R^s_n\,\hat G_{sp}\,[\,\hat G^{-1}]^{kr}.
\endgathered
\mytag{3.14}
$$
As we already mentioned above, passing from $\hat\bold T$ to the deformation
tensor $\hat\bold G$, we loose a part of information contained in $\hat
\bold T$. Nine components of the tensor $\bold R$ in \mythetag{3.14} is an
exact quantitative measure for that loss. We cannot express $\bold R$ 
through $\hat\bold G$ and $\nabla\hat\bold G$, hence, we cannot write the
equation \mythetag{1.4} in terms of the tensor fields $\hat\bold G$ and 
$\bold R$. However, we can do it for some equations derived from
\mythetag{1.4} and \mythetag{1.1}.\par
\head
4. Accessory differential relationships.
\endhead
    Let's consider the equation \mythetag{1.4} again. One can write it in
a formal, but more easily understandable way as follows (see \mycite{1}):
$$
\hskip -2em
\rot\hat\bold T=\boldsymbol\rho.
\mytag{4.1}
$$
Similarly, the equation \mythetag{1.1} is written as
$$
\hskip -2em
\frac{\partial\hat\bold T}{\partial t}+\bold j=-\grad\bold w.
\mytag{4.2}
$$
From \mythetag{4.1} and \mythetag{4.2} one easily derives the following
two relationships:
$$
\gather
\hskip -2em
\divr\boldsymbol\rho=0,
\mytag{4.3}\\
\vspace{2ex}
\hskip -2em
\frac{\partial\boldsymbol\rho}{\partial t}+\rot\bold j=0.
\mytag{4.4}
\endgather
$$
Our present goal in this section is to rewrite \mythetag{4.3} and
\mythetag{4.4} in terms of the real space tensor fields $\hat\bold G$,
$\bold R$, and $\bold J$.\par
    The tensor $\bold R$ is defined by the first formula \mythetag{2.9}.
Its components are expressed through the components of the tensor 
$\boldsymbol\rho$. By means of the inverse matrix $\hat\bold T=\hat
\bold S^{-1}$ one can express $\rho^{\,i}_q$ back through $R^m_q$:
$$
\hskip -2em
\rho^{\,i}_q=\sum^3_{m=1}\hat T^i_m\,R^m_q.
\mytag{4.5}
$$    
The equation \mythetag{4.3} in coordinate form is written as follows:
$$
\hskip -2em
\sum^3_{p=1}\sum^3_{q=1}g^{pq}\,\nabla_{\!p}\rho^{\,i}_q=0.
\mytag{4.6}
$$
Substituting \mythetag{4.5} into \mythetag{4.6}, we immediately derive
the equality
$$
\hskip -2em
\sum^3_{m=1}\sum^3_{p=1}\sum^3_{q=1}g^{pq}\,\hat T^i_m\,
\nabla_{\!p}R^m_q+\sum^3_{m=1}\sum^3_{p=1}\sum^3_{q=1}
g^{pq}\,R^m_q\,\nabla_{\!p}\hat T^i_m=0.
\mytag{4.7}
$$
Multiplying \mythetag{4.7} by $\hat S^r_{i}$ and summing up on 
the index $i$, we get
$$
\hskip -2em
\sum^3_{p=1}\sum^3_{q=1}g^{pq}\,\nabla_{\!p}R^r_q
+\sum^3_{m=1}\sum^3_{p=1}\sum^3_{q=1}
g^{pq}\,R^m_q\,\hat Z^r_{pm}=0.
\mytag{4.8}
$$
If we remember that $\hat Z^r_{pm}$ can be obtained from the formula
\mythetag{3.14}, we see that \mythetag{4.8} is a differential
equation written in terms of the tensor fields $\bold R$ and $\hat
\bold G$. This equation is a purely real space substitute for
\mythetag{4.3}.\par
     In order to perform the same transformations with the equation
\mythetag{4.4}, let's apply the time derivative $\partial/\partial t$
to \mythetag{4.5}. As a result we get
$$
\hskip -2em
\sum^3_{k=1}\frac{\partial\hat T^i_k}{\partial t}\,R^k_q+
\sum^3_{k=1}\hat T^i_k\,\frac{\partial R^k_q}{\partial t}+
\sum^3_{m=1}\sum^3_{r=1}\sum^3_{s=1}g_{qm}\,
\omega^{mrs}\,\nabla_{\!r}j^{\,i}_{\,s}=0.
\mytag{4.9}
$$
For $j^{\,i}_{\,s}$ in \mythetag{4.9} there is a formula similar to \mythetag{4.5}. It is derived from \mythetag{2.9}:
$$
\hskip -2em
j^{\,i}_s=\sum^3_{m=1}\hat T^i_m\,J^m_s.
\mytag{4.10}
$$    
Substituting \mythetag{4.10} into \mythetag{4.9} and applying 
\mythetag{1.1} and \mythetag{1.3}, we obtain
$$
\pagebreak
\gathered
\frac{\partial R^k_q}{\partial t}-\sum^3_{m=1}J^{\,k}_m\,R^m_q
-\sum^3_{m=1}\nabla_{\!m}v^k\,R^m_q-\sum^3_{m=1}\sum^3_{p=1}
v^p\,\hat Z^k_{mp}\,R^m_q\,+\\
+\sum^3_{m=1}\sum^3_{r=1}\sum^3_{s=1}g_{qm}\,\omega^{mrs}\,
\nabla_{\!r}J^{\,k}_{\,s}+
\sum^3_{m=1}\sum^3_{r=1}\sum^3_{s=1}\sum^3_{p=1}g_{qm}\,\omega^{mrs}\,
\hat Z^k_{rp}\,J^{\,p}_{\,s}=0.
\endgathered\quad
\mytag{4.11}
$$
Like \mythetag{4.8}, the equation \mythetag{4.11} is a purely real space
substitute for the equation \mythetag{4.4}. Though \mythetag{4.8} and
\mythetag{4.11} are derived from the accessory equations \mythetag{4.3} 
and \mythetag{4.4}, now they play more substantial role because of the
absence of a real space substitute for the equation \mythetag{4.1}.\par
    The equations \mythetag{4.3} and \mythetag{4.4} are compatible.
This means that one cannot produce new equations of the same or lower order
by differentiating them and combining the resulting equalities. Otherwise,
if that were not the case, the newly produced equations would be called
the {\it compatibility conditions}. Though the equations \mythetag{4.8} 
and \mythetag{4.11} are derived from \mythetag{4.3} and \mythetag{4.4},
their compatibility is not so evident. The study of all possible 
compatibility conditions in the real space version of the theory of
dislocations is the subject for a separate paper. It should be done
with the use of differential geometric methods and terminology.
\head
5. Acknowledgments.
\endhead
    I am grateful to Gregg Allen whose remark during the rehearsal of
my report at The University of Akron in January 2004 stimulated my
interest to strain-stress curves in the material science. The ultimate
goal of the series of papers initiated by \mycite{1} is to calculate
numerically and thus predict the strain-stress curves of crystalline
materials. The same activity for amorphous materials was initiated
in \mycite{4}. 

\enddocument
\end